\documentclass[reprint,superscriptaddress,amsmath,amssymb,aps,prl]{revtex4-2}

\usepackage{graphicx}

\usepackage{color}
\usepackage{bbm}
\usepackage{tensor}
\usepackage[all,cmtip]{xy}
\usepackage{dcolumn}
\usepackage{bm}

\newcommand{\overbar}[1]{\mkern 1.5mu\overline{\mkern-1.5mu#1\mkern-1.5mu}\mkern 1.5mu}
\def\TT{{ \mathrm{T} \overbar{\mathrm{T}}}}

\makeatletter
\newcommand{\raisemath}[1]{\mathpalette{\raisem@th{#1}}}
\newcommand{\raisem@th}[3]{\raisebox{#1}{$#2#3$}}
\makeatother

\usepackage{tensor}
\usepackage{physics}
\usepackage{fontawesome}
\usepackage{slashed}
\usepackage{amsmath,amssymb,calc, amsthm,bbm, epsfig,psfrag, mathtools}
\usepackage[normalem]{ulem} 

\def\cO{{\mathcal{O}}}

\usepackage{latexsym,bm,amsfonts}
\usepackage{graphicx, enumerate}
 \usepackage[all,cmtip]{xy}
  \usepackage{float}
\allowdisplaybreaks

\newcommand{\Comment}[1]{{}}
\definecolor{darkblue}{rgb}{0.15,0.35,0.55}
\definecolor{reddish}{rgb}{0.65, 0.2, 0.2}
\usepackage[linktocpage=true]{hyperref}
\hypersetup{
colorlinks=true,
citecolor=darkblue,
linkcolor=reddish,
urlcolor=darkblue,
pdfauthor={},
pdftitle={},
pdfsubject={}
}

\def\be{\begin{equation}}
\def\ee{\end{equation}}
\def\bea{\begin{eqnarray}}
\def\eea{\end{eqnarray}}

\newfont{\goth}{ygoth.tfm scaled 1200}                   

\def\1{{(1)}}
\def\2{{(2)}}
\def\3{{(3)}}

\newcommand{\geta}{g}
\def\TT{{T\overbar{T}}}

\newcommand {\cA}{{\cal A}}

\newcommand {\cF}{{\cal F}}

\newcommand {\cL}{{\cal L}}

\newcommand {\cN}{{\cal N}}
\newcommand {\cP}{{\cal P}}

\newcommand {\cS}{{\cal S}}
\newcommand {\cT}{{\cal T}}
\newcommand {\cU}{{\cal U}}

\def\d{\delta}

\def\l{\lambda}

\def\L{\Lambda}

\newcommand{\au}{{\underline{a}}}

\newcommand{\pa}{\partial}

\newcommand{\non}{\nonumber}
\newcommand{\ba}{\begin{array}}
\newcommand{\ea}{\end{array}}

\def\double #1{#1{\hbox{\kern-2pt $#1$}}}

\newcommand{\bsubeq}{\begin{subequations}}
\newcommand{\esubeq}{\end{subequations}}

\newcommand{\STr}{\mathrm{STr }}

\def\tr{{\rm tr}}

\begin{document}

\title{ $\TT$-like Flows of Yang-Mills Theories}

\author{Christian Ferko}
\email{c.ferko@northeastern.edu}
\affiliation{Department of Physics, Northeastern University, Boston, MA 02115, USA}
\affiliation{The NSF Institute for Artificial Intelligence
and Fundamental Interactions}

\author{Jue Hou}
\email{juehou@seu.edu.cn}
\affiliation{School of Physics \& Shing-Tung Yau Center,  Southeast University, Nanjing 211189, P. R. China}%

\author{Tommaso Morone}
\email{tommaso.morone@unito.it}
\affiliation{Dipartimento di Fisica, Universit\`a di Torino, and INFN Sezione di Torino,  Via P. Giuria 1, 10125, Torino, Italy}

\author{Gabriele Tartaglino-Mazzucchelli}
\email{g.tartaglino-mazzucchelli@uq.edu.au}
\affiliation{School of Mathematics and Physics, University of Queensland, St Lucia, Brisbane, Queensland 4072, Australia}%

\author{Roberto Tateo}
\email{roberto.tateo@unito.it}
\affiliation{Dipartimento di Fisica, Università di Torino, and INFN Sezione di Torino, Via P. Giuria 1, 10125, Torino, Italy}%

\date{\today}

\begin{abstract}
We study $\TT$-like deformations of $d>2$ Yang-Mills theories. The standard $\TT$ flows lead to multi-trace Lagrangians, and the non-Abelian gauge structures make it challenging to find Lagrangians in a closed form. However, within the geometric approach to $\TT$,  we obtain the closed-form solution to the metric flow and stress-energy tensor, and show that instanton solutions are undeformed. We also introduce new symmetrised single-trace $\TT$-like deformations, whose solutions in $d=4$ include the non-Abelian Born-Infeld Lagrangian proposed by Tseytlin in 1997.
\end{abstract}

\maketitle

\section{Introduction}

$\TT$ deformations, originally formulated as irrelevant deformations of two-dimensional field theories \cite{Cavaglia:2016oda, Smirnov:2016lqw}, have received significant attention due to their remarkable non-perturbative effects and solvable nature. Higher-dimensional generalizations of classical $\TT$-like flows have been proposed in \cite{Taylor:2018xcy, Bonelli:2018kik, Conti:2018jho, Cardy:2018sdv, Ferko:2019oyv, Babaei-Aghbolagh:2020kjg, Conti:2022egv, Hou:2022csf, Ferko:2023sps,Ferko:2023wyi, Ferko:2024zth}. 
Given a one-parameter family of Lagrangians, $\cL=\cL^{(\l)}$, and denoting the associated stress-energy tensor by $T_{\mu\nu}=T^{(\l)}_{\mu\nu}$,
we consider the Lagrangian flow in $d$ space-time dimensions defined by the following partial differential equation:
\be
\frac{\pa\cL^{(\l)}}{\partial\lambda}
=f(T^{(\l)}_{\mu\nu})
=
\frac{1}{2d}\Big(T^{(\l)\mu\nu}T^{(\l)}_{\mu\nu}-\frac{2}{d}(T^{(\l)\mu{}}{}_{\mu})^2\Big)
~ .
\label{flow}
\ee
When $d=2$, 
this is proportional to ${\rm det}\, T_{\mu\nu}$, which is the $\TT$ operator introduced by Zamolodchikov in \cite{Zamolodchikov:2004ce}.
We will denote \eqref{flow} as the double-trace $\TT$ flow equation. 

This paper addresses the question of finding closed-form solutions for $\TT$-like flows for non-Abelian Yang-Mills theories in arbitrary space-time dimensions.

While stress-energy tensor deformations of Abelian gauge theories, leading to the emergence of Born-Infeld functionals (BI) in $d=4$, have been extensively discussed in the literature \cite{Conti:2018tca, Ferko:2021loo,Ferko:2022iru}, the complexity of the PDEs emerging in the non-Abelian scenario (even for low-rank gauge groups) dramatically grows as the space-time dimensions increase, with significant simplifications arising only in the two-dimensional setting. Nevertheless, the existence of auxiliary metric flows associated with $\TT$-like deformations 
makes it possible to directly compute the deformed Hamiltonian density $\mathcal{H}^{(\l)}$, which also 
satisfies \eqref{flow}, up to a minus sign. 

We also observe that the non-linear
gauge theories obtained from the $\TT$-like deformation of the Yang-Mills Lagrangian differ from the several proposals for possible non-Abelian extensions discussed in the literature. In this sense, stress tensor deformations can provide a mechanism for generating non-linear solutions through a Lagrangian flow. 
Moreover, we show how a notion of a symmetrized single-trace stress-energy tensor flow can be introduced, which, in $d=4$, reproduces the non-Abelian Born-Infeld Lagrangian proposed by Tseytlin \cite{Tseytlin:1997csa}.
\section{double-trace $\TT$ flows}
\label{sec:double_trace}
In this section, we review some key results previously obtained in the literature regarding $2d$ and $4d$ $\TT$ flows, that play a central role in our analysis. Then, we present a new result for the
$\TT$ flow \eqref{flow} for Yang-Mills in $d=4$.

The $\TT$ deformation of $2d$ Yang-Mills was studied in \cite{Conti:2018jho,Ireland:2019vvj,Brennan:2019azg}. Thanks to  properties of the $2d$ $\TT$ operator,
this deformation might represent the first example of a quantum mechanically well-defined BI-type extension of a Yang-Mills theory in $d=2$. Given a set of generators $T_{\underline{a}}$ in a representation of a compact gauge group $G$, satisfying $[T_{\underline{a}},T_{\underline{b}}]=f_{\underline{a}\underline{b}}{}^{\underline{c}}T_{\underline{c}}$ and $\Tr[T_{\underline{a}} T_{\underline{b}}]=\d_{\underline{a}\underline{b}}$, the field strength $\mathbb{F}_{\mu\nu}=T_{\underline{a}}\cF^{\underline{a}}_{\mu\nu}$ is defined in terms of the gauge connection $\mathbb{A}_\mu=T_{\underline{a}}\cA^{\underline{a}}_{\mu}$ and the gauge coupling $g$ as
\be
\cF^{\underline{a}}_{\mu\nu}
=\pa_{\mu}\cA^{\underline{a}}_{\nu}
-\pa_{\nu}\cA^{\underline{a}}_{\mu}
+gf_{\underline{b}\underline{c}}{}^{\underline{a}}\cA^{\underline{b}}_\mu \cA^{\underline{c}}_\nu
~.
\label{Fmunu-1}
\ee
The undeformed Yang-Mills Lagrangian is \footnote{Adjoint gauge indices are raised and lowered with the Cartan-Killing matrix $\Tr[T_{\underline{a}} T_{\underline{b}}]=\d_{\underline{a}\underline{b}}$ and its inverse $\d^{\underline{a}\underline{b}}$. In the paper, we will denote as $\mathbb{Id}$ the identity matrix acting in the vector space of the $T_{\underline{a}}$ representation.}
\be
\cL_{\rm YM}
=-\frac{1}{4}\Tr[\mathbb{F}^{\mu\nu}\mathbb{F}_{\mu\nu}]
=-\frac{1}{4}\cF_{\underline{a}}^{\mu\nu}\cF^{\underline{a}}_{\mu\nu}
=\frac{1}{4}x_1
~,
\label{LYM-0}
\ee
where we have introduced the notations
\be
x_n
:=
\tr[X^n]
~,~~
X\equiv X_\mu{}^\nu
:=
\Tr[
\mathbb{F}_{\mu\rho}
\mathbb{F}^{\rho\nu}
]
=
\cF^{\underline{a}}_{\mu\rho}\cF_{\underline{a}}^{\rho\nu}
~,
\label{xX}
\ee
and the trace, $\tr$, is taken over Lorentz indices, e.g. $x_1= X_\mu{}^\mu$, $x_2= X_\mu{}^\nu X_\nu{}^\mu$, etc. 
In $d=2$, $X_\mu{}^\nu =\frac{1}{2}\d^\mu_\nu x_1$ and an appropriate Ansatz to solve  \eqref{flow} with initial condition $\cL^{(0)}=\cL_{\rm YM}$ is $\cL=\cL^{(\l)}(x_1)$. 
Then \eqref{flow} turns into
\be
\frac{\pa\cL}{\pa\l}
=-\frac{1}{2}\cL^2
+2x_1\cL\frac{\pa\cL}{\pa x_1}-2x_1^2\left(\frac{\pa\cL}{\pa x_1}\right)^2
~,
\label{PDE-2d-YM}
\ee
which is solved by a hypergeometric function \cite{Conti:2018jho,Brennan:2019azg}.
The exact solution of the $d=2$ flow equation \eqref{PDE-2d-YM} does not see whether the undeformed theory is Abelian or non-Abelian, as the only functional dependence is upon $x_1$.

In $d=4$, one of the remarkable discoveries in the study of $\TT$-like flows has been the realisation that the Abelian Born-Infeld theory of non-linear electrodynamics,
\bea
    \mathcal{L}_{\rm{BI}} = \frac{1}{\l}\Bigg\{
    ~1
    -\sqrt{1
    -\frac{1}{2}\l x_1
    +\frac{1}{8}\l^2(x_1^2-2x_2)}
    ~\Bigg\}
    ~,~~~~~~
    \label{gBI-0}
\eea
satisfies eq.~\eqref{flow} with initial condition $\cL_{\rm Maxwell}\!=\!\frac{1}{4}x_1\!=\!-\frac{1}{4}\cF^{\mu\nu}\cF_{\mu\nu}$ and $d\!=\!4$ \cite{Conti:2018jho}. This result has been generalised in various directions. 
The $4d$ $\TT$ deformation of the family of classically conformal and electro-magnetic duality invariant electrodynamics described by the ModMax theory \cite{Bandos:2020hgy}, was lifted to the ModMax-BI Lagrangian which, besides \eqref{flow}, also satisfies a classically marginal root-$\TT$ flow \cite{Ferko:2022iru,Babaei-Aghbolagh:2022uij,Ferko:2022cix,Conti:2022egv}.
As in the $2d$ case, the $\TT$-like flow equations for the deformation of $4d$ free Maxwell turn into partial differential equations that are similar to eq.~\eqref{PDE-2d-YM} \cite{Ferko:2021loo,Ferko:2022iru}.
Exact solutions exist since, in the Abelian case, both $X^\mu{}_\nu$ and $T^{\mu\nu}$ have two eigenvalues with double degeneracy. Hence, any $\TT$-like deformation of Abelian gauge theories based on $\cF_{\mu\nu}$ alone depends only on $x_1$ and $x_2$.
Given these results, some natural questions arise. Can we solve the $\TT$ flow equation \eqref{flow} in $d=4$ for a non-Abelian gauge theory? If so, how is the solution related to non-Abelian extensions of 
Born-Infeld? In this section, we will start to elaborate on the first question.

We aim at finding the solution of \eqref{flow} with initial condition  $\cL^{(0)}=\cL_{\rm YM}$ in $d>2$. 
Some features of the $2d$ and Abelian $4d$ cases remain true, though the algebraic complexity of the problem substantially increases. 
An Ansatz for solving the flow equation \eqref{flow} is given by a Lagrangian $\cL^{(\l)}=\cL^{(\l)}(X_{\mu\nu\rho\tau})$, 
with $X_{\mu\nu\rho\tau}$ given by 
\bsubeq
\bea
&X_{\mu\nu\rho\tau}
:=
\Tr[\mathbb{F}_{\mu(\rho}\mathbb{F}_{\tau)\nu}]
=\cF_{\underline{a}}{}_{\mu(\rho}\cF^{\underline{a}}_{\tau)\nu}
~,
\\
&
X_{\mu\nu\rho\tau}
=X_{(\mu\nu)(\rho\tau)}
=X_{\rho\tau\mu\nu}
~,
\\
&
X_{\mu\nu}
=\eta^{\nu\rho}X_{\mu\nu\rho\tau}
=-2\eta^{\mu\nu}X_{\mu\nu\rho\tau}
~.
\eea
\esubeq
The emergence of this tensor can be explained as follows: when computing the Hilbert energy-momentum tensor to evaluate the right-hand side of \eqref{flow}, derivatives of Lorentz invariants constructed out of $X_{\mu\nu\rho\tau}$, e.g. $x_n$ of \eqref{xX}, with respect to the metric lead to tensors constructed out of $X_{\mu\nu\rho\tau}$ itself.
Upon evaluating $T^\mu{}_{\mu}$ and $T^{\mu\nu}T_{\mu\nu}$, any composite Lorentz invariant based on contracted products of $X_{\mu\nu\rho\tau}$ could emerge.
The classification of these invariants for $d>2$ is beyond the scope of this letter. Here, it suffices to stress that the complexity of the PDEs for $\cL^{(\l)}(X_{\mu\nu\rho\tau})=\cL^{(\l)}(x_1,x_2,\cdots)$ dramatically grows as $d$ increases. 
For this reason, in general $d>2$, no closed-form solution to \eqref{flow} is known. However, using Mathematica codes, solving  eq.~\eqref{flow} to a fixed order in $\l$ is straightforward.  
For $d=4$, up to order $\l^3$ we obtain
\begin{align}\label{LYM-4d}
    \mathcal{L} &= \frac{1}{4} x_1 + \frac{\lambda}{32} \left( 4 x_2 - x_1^2 \right) \nonumber \\
    &\quad + \frac{\lambda^2}{128} \left( x_1^3 - 8 x_1 x_2 + 8 x_3 + 8 X^{\mu \nu} \tensor{X}{^\rho_\mu_\nu^\sigma} X_{\rho \sigma} \right) \nonumber \\
    &\quad + \frac{\lambda^3}{6144} \Bigg( 31 x_1^4 - 424 x_1^2 x_2 + 640 x_1 x_3 + 336 x_2^2  \nonumber \\    
    &\qquad \qquad  - 320 x_4 + 256 x_1 X^{\mu \nu} \tensor{X}{^\rho_\mu_\nu^\sigma} X_{\rho \sigma} \nonumber \\
    &\qquad \qquad + 198  \tensor{X}{^\rho_\mu_\nu^\sigma} X_{\rho \sigma} X^{\alpha \mu \nu \beta} X_{\alpha \beta} \Bigg) + \mathcal{O} ( \lambda^4 ) \, .
\end{align}

Interestingly, if one requires the flow to pass through the $\cL_{\rm YM}$, 
$4d$ Yang-Mills point, then an alternative representation exists.
A consistent Ansatz to solve \eqref{flow} with $\cL^{(0)}=\cL_{\rm YM}$, can be proven to be $\cL^{(\l)}=\cL^{(\l)}(\cS,\cP)$
where  
$\cS=\cS^{\underline{a}}{}_{\underline{b}}
:=\cF^{\underline{a}}_{\mu\nu}\cF_{\underline{b}}^{\mu\nu}$
and $\cP=\cP^{\underline{a}}{}_{\underline{b}}:=\cF^{\underline{a}}_{\mu\nu}\tilde{\cF}_{\underline{b}}^{\mu\nu}$, with 
$\tilde{\cF}^{\underline{a}}_{\mu\nu}=\frac{1}{2}\epsilon_{\mu\nu\rho\tau}\cF^{\underline{a}\,\rho\tau}$. 
In this case, Lorentz invariants based on $X_{\mu\nu\rho\tau}$ can be classified by gauge invariant combinations of $\cS$ and $\cP$.
For a general gauge group, and after moving to Euclidean signature, which is useful for our later discussion of instantons, eq.~\eqref{LYM-4d} turns into 
\begin{align}
\mathcal{L}^{(\l)}=&
-\frac{1}{4}{\mathfrak{Tr}}[\cS]
+\frac{\lambda}{2^5}{\mathfrak{Tr}}\left[\cS^2\!-\cP^2\right]
+\frac{\lambda ^2}{2^7}{\mathfrak{Tr}}
\left[
\cS(\cP^2\!-\cS^2)
\right]
\nonumber\\
&+\frac{1}{3}\frac{ \lambda^3}{2^{11}}\Big(
4 {\mathfrak{Tr}}\left[(4\cS^2\!-\cP^2)(\cS^2\!-\cP^2)
+\cS[\cP,\cS]\cP\right]
\nonumber\\
&~~~~~~~~~~~
-\left({\mathfrak{Tr}}[\cP^2\!-\cS^2]\right)^2
\Big)
+\cO(\lambda^4)
~,
\label{LYM-4d-2}
\end{align}
where $\mathfrak{Tr}$ denotes the trace over adjoint indices.

\section{$4d$ Non-Abelian Born-Infeld theories}
We have just described a non-Abelian extension of the Born-Infeld theory based on the solution of the double-trace flow \eqref{flow}. 
Let us compare it with previous proposals for $4d$, non-Abelian Born-Infeld theories.

A natural proposal is given by (see for example \cite{Argyres:1989qr})
\be
\cL_{\rm Tr-BI}
=
\frac{1}{\alpha^2}\Tr\Big{[}\,\mathbb{Id}-\sqrt{-{\rm det}\big{[}\eta_{\mu\nu}\mathbb{Id}+\alpha \mathbb{F}_{\mu\nu}\big{]}}\Big{]}
~,
\label{LBI-AN}
\ee
where the determinant is over the Lorentz indices.
For a generic gauge group, this cannot coincide with \eqref{LYM-4d} since the latter is multitrace, while \eqref{LBI-AN} is single-trace.

A prescription to constrain a non-Abelian BI theory is to match it with open-string calculations. It is known that the $F^4$ contribution to the open string effective action takes the following special form
\be
\STr\Big{[}
\tr{\,\mathbb{F}^4}
-\frac{1}{4}\big(\tr{\,\mathbb{F}^2}\big)^2
\Big{]}
~.
\label{STr-F4}
\ee
Given a tensor $\cU^{\underline{a}_1\cdots\underline{a}_p}$, which is associated to the matrix
$\mathbb{U}
=
T_{\underline{a}_1}
\cdots 
T_{\underline{a}_p} \cU^{\underline{a}_1\cdots\underline{a}_p}$, 
its symmetrized trace ($\STr$) is 
\bea\label{def-STr}
\STr[\mathbb{U}]
:=
\Tr[\underline{\mathbb{U}}]
~&,&~~~~~
{\underline{\mathbb{U}}}
:=
T_{\underline{A}_p}
\cU^{\underline{A}_p}
 ~,
\eea
where we have introduced a multi-index $\underline{A}_n$ which denotes a set of completely symmetrized adjoint indices $\underline{A}_n\!=\!(\underline{a}_1\!\cdots\underline{a}_n)$ and also the contractions ${\underline{\cU}}^{\underline{A}_p}
:=
{{\cU}}^{(\underline{a}_1\!\cdots\!\underline{a}_p)}$
and 
$T_{\underline{A}_p}
:=
T_{(\underline{a}_1}
\!\cdots
T_{\underline{a}_p)}$
\footnote{Symmetrization of $n$ indices includes a $1/n!$ factor.}.
We have also introduced underlined matrices constructed out of completely symmetrized products of generators. Due to \eqref{STr-F4}, 
in \cite{Tseytlin:1997csa}, see also \cite{Bergshoeff:2001dc}, Tseytlin argued that the effective Lagrangian describing the tree-level open strings might be a generalisation of \eqref{LBI-AN} based on a symmetrized single-trace
\be
\cL_{\rm STr-BI}
=
\frac{1}{\alpha^2}\STr\Big{[}\,\mathbb{Id}-\sqrt{-{\rm det}\big{[}\eta_{\mu\nu}\mathbb{Id}+\alpha \mathbb{F}_{\mu\nu}\big{]}}\Big{]}
~.
\label{LBI-Tseytlin}
\ee
Although the $\STr$ prescription fails at orders higher than $\mathbb{F}^4$, see e.g.~\cite{Koerber:2004ze},
this model possesses interesting properties. 
First, the $\STr$  simplifies the Lagrangian, making it effectively Abelian; later we will elaborate more on this point.
Moreover, the $\STr$ might encode a universal sector, as it eliminates the ambiguity in the definition of non-Abelian derivative expansions \cite{Tseytlin:1997csa}.
Coming back to our discussion,
considering its single-trace nature, it is however evident that eq.~\eqref{LYM-4d} cannot match with \eqref{LBI-Tseytlin}.

There have been other proposals for non-Abelian BI; see, e.g., \cite{Hagiwara:1981my,Dyadichev:2001su,CiriloLombardo:2005yy}. In all of these, a matching with \eqref{LYM-4d} is impossible due to the gauge indices trace structures.
The only exception is in \cite{CiriloLombardo:2005yy}, which is based on a quartic-root action where the building blocks are $x_1,\,x_2,\,x_3,\,x_4$ of eq.~\eqref{xX}.
However, from the third order in $\lambda$ on, the Lagrangians of \cite{CiriloLombardo:2005yy} and \eqref{LYM-4d} disagree, even when one modifies the second term in \eqref{flow} to $-r(T_\mu{}^\mu)^2$ for a real constant $r$.

To summarise, the solution of the flow \eqref{flow} deforming pure Yang-Mills, eq.~\eqref{LYM-4d}, is a new type of non-Abelian 
Born-Infeld. Moreover, as far as we could see, none of the previous proposals is a double-trace $\TT$ deformation. However, we will later discuss how \eqref{LBI-Tseytlin} solves a new type of symmetrised single-trace $\TT$-like flow.

\section{Instantons and monopoles}
The algebraic complexity of non-Abelian gauge theories makes it highly non-trivial 
to find a closed form for the double trace $\TT$-deformed Yang-Mills Lagrangian. In this section, we discuss how, despite such difficulties, it is still possible to obtain exact deformed solutions associated with known field configurations of the undeformed theory. We will focus on instantons and monopoles, representing key solutions of $d=4$ Yang-Mills theories.

Instantons are finite-action 
solutions of the Euclidean Yang-Mills theory, 
defined by  the first-order PDE
given by the (anti-)self-duality condition on the gauge field strength, $\tilde{\cF}^{\underline{a}}_{\mu\nu}=\pm\cF^{\au}_{\mu\nu}$.  A 
key property for our analysis is that the Euclidean stress-energy tensor evaluated on an (anti-)instanton 
is identically zero.
Due to their importance, it is natural to seek 
instantons in deformed Yang-Mills theories. For various non-Abelian Born-Infeld 
models, instantons have been identified; see, e.g., \cite{Brecher:1998tv, Park:1999gd}. The symmetrized single-trace action \eqref{LBI-Tseytlin} 
admits undeformed instanton solutions. A priori, there is no guarantee that a BI-type extension of Yang-Mills should admit such configurations.
Does the Euclidean Lagrangian \eqref{LYM-4d-2}, admit them? Intriguingly, yes! And the statement generalises to analogue solutions of theories satisfying the $\TT$ flow \eqref{flow} in any space-time dimension and signature. 

It has been shown in \cite{Conti:2022egv} that the equations of motion (EOM) for the deformed theory $\cL^{(\lambda)}$ can be obtained from the Euler-Lagrange equations of the associated undeformed 
$\cL^{(0)}$ in terms of a field-dependent metric deformation. For generality, here we 
consider an arbitrary background metric tensor $g_{\mu\nu}$. Further, as in \cite{Conti:2022egv}, 
we consider the Lagrangian flows \eqref{flow} with a modified operator
$f(T^{(\l)}_{\mu\nu})
:=
\frac{1}{2d}\Big(T^{(\l)\mu\nu}T^{(\l)}_{\mu\nu}-r(T^{(\l)\mu{}}{}_{\mu})^2\Big)$
with an extra real parameter $r$.
Then, denoting the deformed metric by $h^{(\lambda)}_{\mu\nu}$, such that $h_{\mu\nu}^{(0)} = g_{\mu\nu}$, one has
\be\label{metric_flow}
\frac{d h_{\mu\nu}^{(\lambda)}}{d \l} = \frac{2}{d}\left(T^{(\lambda)}_{\mu\nu} - r T^{(\l)\alpha}{}_\alpha h_{\mu\nu}^{(\lambda)}\right)\,.
\ee
While equation (\ref{metric_flow}) can be solved recursively in $\lambda$ \cite{Conti:2022egv}, one of our main results is that it can also be integrated exactly using the method of characteristics introduced in \cite{Hou:2022csf}.
The exact information on the auxiliary metric $h_{\mu\nu}^{(\lambda)}$ in terms of the initial data can be conveniently encoded in the matrix $\bm{\Omega}:= \Omega^\mu{}_\nu = g^{\mu\alpha}h^{(\l)}_{\alpha\nu}$ (See the supplemental material for more details.):
\bsubeq
\bea
&\bm{\Omega}(\l) 
= 
 \L_1\exp \left(\L_2 (d\, \bm{T}^{(0)} - t_1 \bm{I}) \right)
~,~~~~~~
 \\
&\L_1:=\Big( 1+\frac{4 \l}{d} (d r-1) \left(t_1 \left(\lambda  r t_1-4\right)-\lambda  t_2\right)\Big)^{\frac{2}{d}} 
 ~,~~~~~~
 \\
&\L_2
:=
\frac{4}{d \sqrt{d r - 1} \sqrt{t_1^2 - d t_2}}
\Bigg(\arctan\left(\frac{t_1 \sqrt{d r - 1}}{\sqrt{t_1^2 - d t_2}}\right) 
~~~~~~~~~~
\non\\
&~~~~~~~~~~~~~~~
+ \arctan\left(\frac{\sqrt{d r-1} \left(t_1 \left(\lambda  r t_1-2\right)-\lambda  t_2\right)}{2 \sqrt{t_1^2-d t_2}}\right)\Bigg)
~,~~~~~~~~~
\eea
\label{eq:exact-metric}
\esubeq
$\!\!$where $\bm{T}^{(0)}$ denotes the unperturbed stress-energy tensor $T^{(0)\mu}{}_{\nu}$, $t_n = \tr[(\bm{T}^{(0)})^n]$, and $\bm{I}$ is the identity matrix. It is straightforward to show that when $T_{\mu\nu}^{(0)}$ identically vanishes, the resulting auxiliary metric $h_{\mu\nu}^{(\l)}$ coincides with $g_{\mu\nu}$, thereby ensuring that the associated field configurations remain undeformed along the flow.  This is exactly the case for instanton solutions of the Yang-Mills theory (where $g_{\mu\nu} = \d_{\mu\nu}$), which are therefore preserved by the deformation. The Lagrangian evaluated on solutions with $T_{\mu\nu}^{(0)}=0$ is also undeformed.
A similar approach can be used in the case of monopoles, which are localized solutions to the Yang-Mills equations of motion characterized by magnetic charge. As an illustrative example, we shall focus on Wu-Yang-type monopoles \cite{Mark1970}, which arise in the context of pure $SU(2)$ gauge theories in $d=4$. The associated connection $\mathcal{A}_\mu^{\underline{a}}$ is
\be\label{wy_A}
\mathcal{A}_\mu^{\underline{a}} = Q_m\epsilon_{\mu}{}^{\underline{a}}{}_{\underline{b}}\frac{\hat{R}^{\underline{b}}}{R}\,,
\ee
where $Q_m$ denotes the magnetic charge of the monopole, $\hat{R}$ is a radial versor and $\epsilon$ is the Levi-Civita tensor. In polar coordinates, the stress-energy tensor is
\be
T^{(0)\mu}{}_\nu = -\frac{Q^2_m}{2R^4}\operatorname{diag}(1,1,-1,-1)\,.
\label{eq:TM}
\ee
Setting $g_{\mu\nu}=\eta_{\mu\nu}$, $r=1/2$, and $d=4$ in (\ref{eq:exact-metric}), using (\ref{eq:TM}) and (\ref{metric_flow}), we find
\begin{equation}\label{invert_this}
    T_{\mu\nu}^{(0)} = T_{\mu\nu}^{(\l)}-\frac{1}{2}T^{(\l)\alpha}{}_\alpha h_{\mu\nu}^{(\l)}\,.
\end{equation}
Equation \eqref{invert_this} can be inverted to obtain the deformed stress-energy tensor $T_{\mu\nu}^{(\l)}$ as a function of $T_{\mu\nu}^{(0)}$. 
The Hamiltonian density of the deformed field configuration is then (See the supplemental material.):
\begin{equation}\label{energy density deformed}
   \mathcal{H}^{(\l)}(R) = -T^{(\l)0}{}_0 = \frac{1}{\l}\left(1-\sqrt{1-\frac{\lambda  Q_m^2}{R^4}}\right).
\end{equation}
It is worth noting that the energy distribution \eqref{energy density deformed} is formally identical to that of a point-charge configuration in Born-Infeld electrodynamics. Moreover, one has
\begin{equation}
 \pdv{\mathcal{H}}{\l} =- \frac{1}{8}  \Big(T^{(\l)\mu\nu}T^{(\l)}_{\mu\nu}-\frac{1}{2}(T^{(\l)\mu{}}{}_{\mu})^2\Big) = -\frac{1}{2} \pdv{(R^4 \mathcal{H}^2)}{R^4} . 
\end{equation}
While the Wu-Yang monopole provides a simple toy model in the context of non-Abelian field theories, similar methods can be employed to study the deformed version of other solutions to the Yang-Mills equations of motion, even beyond pure-gauge 
(such as for 't Hooft-Polyakov monopoles \cite{tHooft:1974kcl, Polyakov:1974ek}). 
Although finding the exact form of the deformed Lagrangian remains an open question, the previous results indicate that the dynamics associated with $\TT$-like deformations of Yang-Mills theories are,  to some extent, solvable due to the exact knowledge of the deformed metric (\ref{eq:exact-metric}).

\section{Symmetrized single-trace $\TT$}

We start by describing a one-to-one correspondence between symmetrized single-trace non-Abelian Yang-Mills theories and Abelian theories of non-linear electrodynamics.
Consider the non-Abelian field strength  $\cF^{\underline{a}}_{\mu\nu}$. Given a set of parameters $t_{\underline{a}}$, we introduce $\cF_{\mu\nu}(t)$ as
\be
\cF_{\mu\nu}(t):=t_{\underline{a}}\cF^{\underline{a}}_{\mu\nu}
~,~~
\mathbb{F}_{\mu\nu}=T_{\underline{a}}\pa^{\underline{a}}\cF_{\mu\nu}(t)
~,~~
\pa^{\underline{a}}
=\frac{\pa}{\pa t_{\underline{a}}}
~.
\ee
Given a tensor $\cU^{\underline{A}_n}$, see \eqref{def-STr}, we also introduce a homogeneous polynomial of degree $n$, $\cU(t):=t_{\underline{A}_n}\cU^{\underline{A}_n}$, as well as
$t_{\underline{A}_n}:=t_{\underline{a}_1}\!\cdots t_{\underline{a}_n}$
and $\pa^{\underline{A}_n}:=\pa^{\au_1}\!\cdots\pa^{\au_n}$. 
Then \footnote{We normalise $(T_{\au}\pa^{\au})^0=\mathbb{Id}$}
\bea
{\underline{\mathbb{U}}}
\!=\!
\frac{1}{n!}(T_{\au}\pa^{\au})^n\cU(t)
~,~~
\STr[{\underline{\mathbb{U}}}]
\!=\!\frac{1}{n!}\!\Tr[T_{\underline{A}_n}]\pa^{\underline{A}_n}
\cU(t)
\,.~~~~~\,
\eea
Given a matrix function $\Gamma[\mathbb{U}]$,
an efficient way to represent its projection to the symmetric tensor product of adjoint representations and the symmetrized trace is 
\bsubeq
\bea
&
\underline{\big{[}
\Gamma[\mathbb{U}]
\big{]}}
=
{\mathbb{S}}_{\raisemath{1pt}{\mathbbm{ym}}}
\Gamma[\,\mathbb{\cU}(t)]
~,~~~
{\mathbb{S}}_{\raisemath{1pt}{\mathbbm{ym}}}:=\exp{T_{\au}\pa^{\au}}
~,~~~
\\
&\STr\Big[\Gamma[{\mathbb{U}}]\Big]
=
\Tr\Big{[}\,{\mathbb{S}}_{\raisemath{1pt}{\mathbbm{ym}}}\Gamma[\,\cU(t)]\Big{]}
~.
\eea
\esubeq
We assume $\Gamma(u)$ analytic around $u\!=\!t_\au\!=\!0$, and we take $t_\au\!\equiv\! 0$ after applying the ${\mathbb{S}}_{\raisemath{1pt}{\mathbbm{ym}}}$ operator. The idea of ${\mathbb{S}}_{\raisemath{1pt}{\mathbbm{ym}}}$ is to take the Taylor expansion around $t_\au\!=\!0$ of any function and turn it into a matrix-valued Taylor expansion where $t_{\au_1}\!\cdots t_{\au_n}$, are traded for $T_{(\au_1}\!\cdots T_{\au_n)}$, the building block of the underlined matrices and the symmetrized trace.

Now, given a Lagrangian $\cL=\cL(\cF_{\mu\nu})$ for a theory of Abelian electrodynamics based on a single field strength $\cF_{\mu\nu}$, 
 assuming  $\cL(\cF_{\mu\nu})$ is analytic around $\cF_{\mu\nu}=0$,
 we can uniquely associate a symmetrized single-trace non-Abelian Yang-Mills theory as follows:
\bea
L(\mathbb{F}_{\mu\nu})
:=
\Tr\Big{[}{\mathbb{S}}_{\raisemath{1pt}{\mathbbm{ym}}}\cL(\cF_{\mu\nu}(t))\Big{]}
=
\STr[\cL(\mathbb{F}_{\mu\nu})]
~.
\label{STr-YM-1}
\eea
Under similar assumptions, the reverse is also true. 
Given a symmetrized single-trace theory, there is a unique associated non-linear Abelian electrodynamics.

If we want to construct $\TT$-like flows starting from \eqref{STr-YM-1}, we first need to compute the energy-momentum tensor. We do that by using the Hilbert prescription
\be
T_{\mu\nu}
=T_{\mu\nu}[L(\mathbb{F}_{\rho\tau})]
=-\frac{2}{\sqrt{-g}}\frac{\d S}{\d g^{\mu\nu}}
~.
\label{Hilbert}
\ee
If we plug the Lagrangian \eqref{STr-YM-1} into eq.~\eqref{Hilbert}, $T_{\mu\nu}$ is by construction a gauge invariant, symmetrized single-trace tensor. Hence, any function $f(T_{\mu\nu})$ will necessarily be multitrace. When the seed theory is the undeformed Yang-Mills theory, we have already seen how the solution of the flow \eqref{flow}, eq.~\eqref{LYM-4d}, is highly multitrace. 
To produce closed single-trace deformations, we need to modify the type of operator we are using in the first place.

If we aim to obtain closed flows for symmetrized single-trace Yang-Mills theories, we can adopt a natural, new 
prescription. 
We first define a matrix extension of $T_{\mu\nu}$ as
\be
\mathbb{T}_{\mu\nu}
=
{\mathbb{S}}_{\raisemath{1pt}{\mathbbm{ym}}}\cT_{\mu\nu}(t)
~,
\label{Hilbert-matrix}
\ee
where we have introduced $\cT_{\mu\nu}(t)=T_{\mu\nu}[\cL(\cF_{\rho\tau}(t))]$. Note that $\mathbb{T}_{\mu\nu}$ is gauge covariant and $\Tr[\mathbb{T}_{\mu\nu}]=T_{\mu\nu}$.
Then, we introduce the new symmetrized single-trace $\TT$-like flow defined by the following differential equation:
\be
\frac{\pa L^{(\l)}}{\partial\lambda}
=\Tr\Big{[}
{\mathbb{S}}_{\raisemath{1pt}{\mathbbm{ym}}}f\big(\cT_{\mu\nu}(t),\lambda\big)
\Big{]}
=
\STr\Big{[} f\big(\mathbb{T}_{\mu\nu},\lambda\big)\Big{]}
~ .
\label{flow-NA-STr}
\ee
Here, we have focused on $d=4$, but this construction works in any dimension. For instance, a $d$-dimensional symmetrized single-trace $\TT$ deformation, which extends the double-trace one of \eqref{flow}, is defined by choosing in \eqref{flow-NA-STr} 
\be
f\big(\cT_{\mu\nu}(t)\big)
=\frac{1}{2d}\Big{(}\cT^{\mu\nu}(t)\cT_{\mu\nu}(t)-\frac{2}{d}\big(\cT^\mu{}_{\mu}(t)\big)^2\Big{)}
~.
\label{flow-STr-TTbar}
\ee

When we discussed $4d$ double-trace $\TT$ flows, for example see eq.~\eqref{LYM-4d}, we saw that the key variables were the independent Lorentz invariant combinations constructed out of $X_{\mu\nu\rho\tau}$.
The question is, how many invariants characterise symmetrized single-trace flows? The answer is only two, as for flows of theories based on a single Abelian field strength. The reason is that symmetrized single-trace flows are equivalent to Abelian flows defined in terms of $\cF_{\mu\nu}(t)$. By using the representations given before, it is straightforward to show that, for any gauge group $G$, the following relations hold
\bsubeq
\bea
&\underline{x}_3 
\!=\! \frac{1}{8}\underline{\Big{[}\underline{x}_1\! \left(6\underline{x}_2-\underline{x}^2_1\right)\! \!\Big{]}}
\,,
~~
\underline{x}_4 
\!=\!
\frac{1}{16}\underline{\Big{[}
4\underline{x}^2_2\!+\!4\underline{x}^2_1\underline{x}_2
\!-\!\underline{x}^4_1
\Big{]}}
    \, ,~~~~~~
\\
&\underline{x}_n\!:=\!{\mathbb{S}}_{\raisemath{1pt}{\mathbbm{ym}}}x_n(t)
\!=\!\underline{\Big{[}\tr\big{[}\mathbb{X}^n\big{]}\Big{]}}~,~~~ \mathbb{X}\!:=\!\mathbb{X}_{\mu}{}^{\nu}
\!=\!\mathbb{F}_{\mu\rho}\mathbb{F}^{\rho\nu}
\,.~~~~~~~
\eea
\esubeq
Here $x_n(t)$ is the invariant defined in \eqref{xX} but now expressed in terms of $\cF_{\mu\nu}(t)$. Similar relations can be used to prove that all $\underline{x}_n$ can be expressed in terms of underlined products of $\underline{x}_1$ and $\underline{x}_2$. Once more, this is precisely as in the Abelian case, where any Lagrangian $\cL(\cF_{\mu\nu})$ 
can be expressed as $\cL(\cF_{\mu\nu})=\cL(x_1,x_2)$. 
We then obtain a symmetrized non-abelian extension of these models by simply considering the Lagrangian
\bea
L(\mathbb{F}_{\mu\nu})
:=
{\rm Tr}\Big{[}{\mathbb{S}}_{\raisemath{1pt}{\mathbbm{ym}}}\cL\big(x_1(t),{x}_2(t)\big)\Big{]}
={\rm STr}[\cL(\underline{x}_1,\underline{x}_2)]
~.~~~~~~
\label{sSTrYM-1}
\eea
Remarkably, if $\cL(x_1,x_2)=\cL(x_1,x_2,\lambda)$ depends on some parameter, 
and satisfies a $\TT$-like flow with operator $f\big(T_{\mu\nu},\lambda\big)$, 
then $L(\mathbb{F}_{\mu\nu},\lambda)$ satisfies the flow equation \eqref{flow-NA-STr}. 

To conclude, we can now study the $4d$ symmetrized single-trace $\TT$ flow with the initial condition given by $L^{(0)}=\cL_{\rm YM}$, the pure Yang-Mills action of \eqref{LYM-0}. From the discussion above, it is clear that the solution takes the same form of the Abelian BI theory 
\be
L^{(\lambda)}
\!=\!
\frac{1}{\lambda}{\rm Tr}\Bigg{[}{\mathbb{S}}_{\raisemath{1pt}{\mathbbm{ym}}}\Bigg(
\!1-\sqrt{1\!-\!\frac{\l}{2} x_1
\!+\!\frac{\lambda^2}{8}\big(x_1^2\!-\!2x_2\big)}
\Bigg)
\Bigg{]}
~,
\label{STr-YM-2}
\ee
with $x_1=x_1(t)$ and $x_2=x_2(t)$.
By using symmetrized algebraic relations, which, as described before, are identical to the Abelian case, and by identifying $\lambda=\alpha^2$, one can easily show that \eqref{STr-YM-2} coincides with  \eqref{LBI-Tseytlin}.

Instanton solutions associated with the undeformed Yang-Mills action are preserved along the symmetrized single-trace flow. As in the double-trace case, this feature ultimately originates from the fact that the energy-momentum tensor of instantonic configurations vanishes.

\section{Conclusions and outlook}
We have initiated the study of $\TT$-like deformations of $d>2$ Yang-Mills theories. Our main results are: 1) for double trace $\TT$ deformations, we obtained a new closed-form solution to the metric flow, allowing us to compute exact observables, even without knowing the Lagrangian; 2) we introduced new symmetrised single-trace $\TT$-like deformations, for which exact deformed Lagrangians can be easily found. Several directions for future research emerge from these results. Exploring deformations of $\cN=4$ super Yang-Mills and studying models coupled to gravity \cite{Morone:2024ffm, Babaei-Aghbolagh:2024hti, Brizio:2024arr} are just two of many exciting possibilities.

It is also natural to study symmetrized, single-trace deformations in other dimensions and for different types of models. In $d=2$, it would be remarkable if some of the unique quantum and solvability properties of $\TT$, as well as its relevance for deformed holography, still hold within this new framework. 
Moreover, while the geometrization of double-trace stress tensor flow has been studied in depth, an analogue mechanism for the symmetrized, single-trace is a new territory to explore.

\begin{acknowledgments}
\medskip
\noindent\textbf{Acknowledgements}
We are grateful to Nicol\`o Brizio, Zejun Huang, and Stefano Negro for helpful discussions. 
C.\,F. is supported by the National Science Foundation under Cooperative Agreement PHY-2019786 (the NSF AI Institute for Artificial Intelligence and Fundamental Interactions).
J.\,H. is supported by the China Postdoctoral Science Foundation No.\,2024M750404 and the Jiangsu Funding Program for Excellent Postdoctoral Talent.
G.\,T.-M. has been supported by the Australian Research Council (ARC) Future Fellowship FT180100353, ARC Discovery
Project DP240101409, and the Capacity Building Package of the University of Queensland.
T.\,M. and R.\,T. received partial support from the INFN project ``Statistical Field Theory (SFT),'' and the Prin (Progetti di rilevante interesse nazionale)  Project No. 2022ABPBEY, with the title ``Understanding quantum field theory through its deformations'', funded by the Italian Ministry of University and Research.
C.\,F., G.\,T.-M. and R.\,T. acknowledge support during the MATRIX Program ``New Deformations of Quantum Field and Gravity Theories,'' 
(Creswick, 22 Jan -- 2 Feb 2024). 
C.\,F. and G.\,T.-M. are grateful to the participants of the meeting ``Integrability in low-supersymmetry theories,'' (Trani, 22 July -- 2 Aug 2024), funded by the COST Action CA22113 by INFN and by Salento University, for stimulating discussions.
\end{acknowledgments}

\bibliographystyle{apsrev4-2} 
\bibliography{master}

\newpage

\onecolumngrid

\appendix
\setcounter{section}{1}
\setcounter{equation}{0}
\begin{center}{\large \textbf{SUPPLEMENTAL MATERIAL
\\\vspace{0.1cm}
 (APPENDICES)
}
}
\end{center}
In the Supplemental Material accompanying our letter, we sketch the derivation of the exact expression for the auxiliary metric  $h_{\mu\nu}^{(\lambda)}$, using the method of characteristics, and give more details on the solution of the Hamiltonian density flow equation for the deformed Wu-Yang monopole.
Note that the equations we refer to as ($M\#$) in this Supplemental Material are contained in the main manuscript.

\vspace{0.5cm}
\begin{center}{\large 
\textbf{
A.  Exact Expression for the Deformed Metric}}
\end{center}
\vspace{0.15cm}

In the following, we shall use  Euclidean signature, the same results hold in Lorentz signature but for the replacement $\sqrt{h} \rightarrow \sqrt{-h}$, $\sqrt{\geta} \rightarrow \sqrt{-\geta}$. As shown in \cite{Conti:2022egv}, the equations of motion associated to $\TT$-like deformed field theories in arbitrary dimensions can be obtained in terms of a field-dependent deformed metric $h^{(\l)}_{\mu\nu}$, such that $h^{(0)}_{\mu\nu} = g_{\mu\nu}$, and
\begin{equation}\label{eoms_metric_def}
 \mathrm{EOMs}^{(\l)}[h^{(\l)}] =  \mathrm{EOMs}^{(0)}[g]\,. 
\end{equation}
From the method of characteristics \cite{Hou:2022csf}, we know that the metric flow equation 
\begin{equation}
\label{metric_flow2}
\frac{d h_{\mu\nu}^{(\lambda)}}{d \lambda} 
= \frac{2}{d}\left(T^{(\lambda)}_{\mu\nu} - r T^{(\l)\alpha}{}_\alpha  h_{\mu\nu}^{(\lambda)}\right)\,,
\end{equation}  
can be recast in the compact form 
\begin{equation}
\label{dGdsGA}
\frac{d \bm{h}(s)}{ d s}=\bm{h}(s) \bm{A}(s)\,,\,\,\,\,\,\,
\frac{d \lambda}{d s}=-2\,,
\end{equation}
where $\bm{h}(s)$ is the matrix form of the metric, $\bm{h}(s)=(h_{\mu\nu}^{(s)})$, and we take the solution about $\lambda$ as $\lambda=-2s$.  By equations (6.8), (6.14), and (6.16) in \cite{Hou:2022csf}, the matrix $\bm{A}$ can be expressed as
\begin{equation}\begin{split}
&\bm{A}(s)=\frac{1}{\sqrt{h(s)}} \left( f(s) \bm{I}+ \bm{B} \right)\,,\\
\end{split}
\end{equation}
with
\begin{equation} \label{inidatah0}
\sqrt{h(s)} =(r d-1) \psi  s^2+\frac{2}{d}(r d-1)t_1s \sqrt{\geta}+
\sqrt{\geta}\,, 
\end{equation}
\begin{equation} \label{inidatah1} \begin{split}
&f(s)=\frac{4}{d}(r d-1)\psi s, \,\,\,\,  \psi =\frac{\sqrt{\geta}}{d}(r (t_1)^2-t_2)\,,\\
\end{split}
\end{equation}
and
\begin{equation} \label{inidatah2}
\bm{B} =-\frac{4 \sqrt{\geta}}{d}(\bm{T}^{(0)}-r t_1 \bm{I})\,.
\end{equation}
In (\ref{inidatah2}), the matrix $\bm{T}^{(0)}$ represents the unperturbed stress-energy tensor $T^{(0)\mu}{}_{\nu}$ in matrix form, where $t_n = \tr[(\bm{T}^{(0)})^n]$, and $\bm{I}$ denotes the identity matrix. From these definitions, it follows that the matrix $\bm{A}$ depends solely on $\bm{T}^{(0)}$ and the parameters $r$, $d$, and $s=-\frac{\lambda}{2}$. Moreover, since $[\bm{T}^{(0)}, \bm{I}] = 0$ trivially, it also holds that $[\bm{A}(s_1), \bm{A}(s_2)] = 0$. Consequently, there is no need to be concerned about the ordering when performing a perturbative expansion. The formal solution of \eqref{dGdsGA} is \cite{Hou:2022csf}
\begin{equation}\begin{split}\label{solG}
\bm{h}(s)= \bm{h}(0) \exp \left( \int_0^s ds' \bm{A}(s') \right)\,.
\end{split}
\end{equation}
This matrix also depends only on $\bm{T}^{(0)}$, $r$, $d$, $s$, and the undeformed metric $\bm{h}(0) = (\geta_{\mu\nu})$. When $T^{(0)\mu}{}_{\nu} = 0$, $\bm{A}$ becomes the zero matrix, and $h_{\mu\nu} = \geta_{\mu\nu}$. For a general $\bm{T}^{(0)}$, equation \eqref{solG} can be integrated, yielding the exact solution:
\begin{align}
\bm{h}(s) &= \bm{h}(0) \;  \left( 1+\frac{s}{d}(d r-1)(2 t_1+r s (t_1)^2-s t_2)\right)^{\frac{2 }{d}} \nonumber \\
&\times \exp \left(\frac{4 (d\, \bm{T}^{(0)}-  t_1 \bm{I}) \left(\arctan\left(\frac{t_1 \sqrt{d r-1}}{\sqrt{(t_1)^2-d t_2}}\right)-\arctan\left(\frac{\sqrt{d r-1} \left(r s (t_1)^2- s t_2+ t_1\right)}{\sqrt{(t_1)^2-d t_2}}\right)\right)}{d \sqrt{d r-1} \sqrt{(t_1)^2-d t_2}}\right)\,.
\label{eq:exact}
\end{align}
Notice that $\tr[d \, \bm{T}^{(0)} - t_1 \bm{I}] = 0$. By using the identity $\det(\exp(\ast)) = \exp(\tr(\ast))$, one can verify that the determinant of $\bm{h}(s)$ consistently matches the expression given in \eqref{inidatah0}. Additionally, we have verified that the perturbative expansion around $s = 0$ fully matches the results obtained using iterative methods, and it reduces to known exact expressions in particular cases \cite{Conti:2022egv, Hou:2022csf, Morone:2024ffm}.
Introducing the matrix $\bm{\Omega}(\l) := \Omega^\mu{}_\nu(\l) = \geta^{\mu\alpha}h_{\alpha\nu}^{(\lambda)}$, we recover the result quoted in the main text in equations ($M15a$)--($M15c$).

The first terms in the small-$\lambda$ expansion are,
\begin{equation}\begin{split}
h_{\mu\nu}^{(\lambda)} &=\geta_{\mu\nu}-\frac{2 \lambda}{d}\left(-T_{\mu\nu}^{(0)}+r T^{(0)\alpha}{}_\alpha\geta_{\mu\nu}\right) \\
&+\frac{\lambda^2}{2d^2}\left(4T_{\mu\rho}^{(0)} T^{(0)\rho}{}_\nu-2 (4 r-d r+1)T_{\mu\nu}^{(0)}T^{(0)\alpha}{}_\alpha
+\geta_{\mu\nu}\left((4 r-d r+1)r (T^{(0)\alpha}{}_\alpha)^2-(r d-1) T^{(0)\alpha}{}_\beta T^{(0)\beta}{}_\alpha \right)\right)+O(\lambda^3).
\end{split}
\label{eq:expansion}
\end{equation}
Concerning the Yang-Mills case in $d=4$, with $r=1/2$ discussed in the main text, with  $\tr[ \bm T^{(0)}]=0$, further simplifications significantly reduce the length of the result. In terms of the  matrix $\Omega^\mu{}_\nu(\l)$, we have
\begin{equation}
\bm{\Omega}(\l) =  \sqrt{1-\frac{\lambda^2}{16} t_2}\,\, \exp\left(2 \bm{T}^{(0)}\frac{\tanh^{-1}(\frac{\lambda}{4} \,\sqrt{t_2})}{\sqrt{t_2}} \right)\,.
\end{equation}
Alternatively, using
\be
\tanh^{-1} (x) = \frac{1}{2}\log\left(\frac{1+x}{1-x}\right)\,,
\ee
we have
\be
\bm{\Omega}(\l) = \sqrt{1-\frac{\lambda^2}{16} t_2} \left(\frac{1+\frac{\lambda}{4}\,\sqrt{t_2}}{1-\frac{\lambda}{4}\,\sqrt{t_2}}\right)^{ \frac{\bm{T}^{(0)}}{\sqrt{t_2}}}\, = \left(1+\frac{\lambda}{4}\,\sqrt{t_2} \right)^{\frac{1}{2}\bm{I}+ \frac{\bm{T}^{(0)}}{\sqrt{t_2}}}\left(1-\frac{\lambda}{4}\,\sqrt{t_2}\right)^{\frac{1}{2}\bm{I}-\frac{\bm{T}^{(0)}}{\sqrt{t_2}}}\,.
\ee
Finally, it is important to note that \eqref{eq:exact} involves a matrix exponential. It is well known that $\exp(\bm{M})$ is convergent if $\bm{M}$ is a finite-dimensional matrix.  Consequently, the auxiliary metric $\bm{h}$ defined through equation (\ref{eq:exact}), as well as the related deformation matrix $\bm{\Omega}$ in equations ($M15a$)--($M15c$), are always well-defined.

\vspace{0.5cm}
\begin{center}{\large 
\textbf{
B.  Deformed Energy Density}}
\end{center}
When computing the deformed Hamiltonian density associated with a specific field configuration, the necessary ingredients reduce to the deformed metric $h^{(\l)}_{\mu\nu}$ and the deformed stress-energy tensor $T^{(\l)}_{\mu\nu}$. Both of them can be obtained from the seed theory data. Specifically, $h^{(\l)}_{\mu\nu}$ is computed by acting on the (undeformed) background metric $g_{\mu\nu}$ with the deformation matrix $\bm{\Omega}$ given by equations ($M15a$)--($M15c$). As far as the deformed stress-energy tensor, there exists a simple procedure that allows the recovery of its form using the identity \eqref{eoms_metric_def}. Let us introduce the auxiliary tensor
\begin{equation}\label{that}
    \widehat{T}^{(\l)}_{\mu\nu}:= - {T}^{(\l)}_{\mu\nu} + r(h^{-1})^{(\l)\alpha\beta}{T}^{(\l)}_{\alpha\beta}{h}^{(\l)}_{\mu\nu}\,,
\end{equation}
which allows us to write $(M14)$ as
\begin{equation}
    \frac{d h_{\mu \nu}^{(\lambda)}}{d \lambda}=-\frac{2}{d} \widehat{T}^{(\l)}_{\mu\nu}\,.
\end{equation}
On the other hand, note that $(M15.a)$ gives us the full form of $h^{(\l)}_{\mu\nu}$ in terms of the seed theory data. It is then sufficient to differentiate $\Omega_{\mu}{}^{\alpha}g_{\alpha\nu}$ with respect to $\lambda$ to obtain $ \widehat{T}^{(\l)}_{\mu\nu}$ as a function of $\l$, $T_{\mu\nu}^{(0)}$ and $g_{\mu\nu}$. Then, \eqref{that} can be easily inverted, yielding
\begin{equation}
 T^{(\l)}_{\mu\nu}= - \widehat{T}^{(\l)}_{\mu\nu} + \frac{1}{dr-1}(h^{-1})^{(\l)\alpha\beta}\widehat{T}^{(\l)}_{\alpha\beta}{h}^{(\l)}_{\mu\nu}\,.   
\end{equation}
There is, however, a hurdle to overcome. In the metric approach, we want to express $T^{(\l)}_{\mu\nu} = T^{(\l)}_{\mu\nu}[h]$ as a function of the deformed auxiliary metric $h^{(\l)}_{\mu\nu}$. 
So far, the current procedure only allows us to express $T^{(\lambda)}_{\mu\nu}$ in terms of the metric $g_{\mu\nu}$. However, significant algebraic complications arise when attempting to invert the relation between the two metric tensors. While numerical techniques can certainly be employed for this purpose, there are special cases in which these problems can be analytically overcome, and the relation between $g_{\mu\nu}$ and $h^{(\lambda)}_{\mu\nu}$ translates into a local deformation of some parameter defining the original metric.\\

As an example, let us consider the Wu-Yang magnetic monopole, with gauge potential defined as in $(M16)$. Due to the spherical symmetry of the configuration, it is convenient to fix a new coordinate system such that the background metric reads
\begin{equation}\label{g_met:radial}
    g_{\mu\nu}dx^\mu dx^\nu = -dt^2 + dR^2 + R^2(d\theta^2+\sin ^2\theta d\phi^2)\,.
\end{equation}
The deformed metric $ h^{(\l)}_{\mu\nu}$ is obtained as 
\begin{equation}
    h^{(\l)}_{\mu\nu} = \Omega_\mu{}^\alpha g_{\alpha\nu}\,, \quad \Omega^{\mu}{}_\nu = \delta^\mu_{\nu} + \frac{\l}{2}T^{(0)\mu}{}_\nu\,.
\end{equation}
In particular, we see that the radial function $R$ in \eqref{g_met:radial} gets deformed into
\begin{equation}\label{z(r)}
\left(1+\frac{\l Q_m^2}{4 R^4}\right)R^2 = R^2 +\frac{\l Q_m^2}{4R^2}:= Z^2\,.   
\end{equation}
From this point of view, $Z^2$ plays the role of an auxiliary radial coordinate for the metric $ h^{(\l)}_{\mu\nu}$, which explicitly depends on the flow parameter $\l$. Inverting equation \eqref{z(r)}, we get
\begin{equation}\label{rofz}
    R^2(Z) = \frac{1}{2}\left(Z^2 + \sqrt{Z^4 - \l Q_m^2}\right)\,.
\end{equation}
Note that, from equation $(M18)$, we have
\begin{equation}
T^{(\l)0}{}_0(R)=  -   \frac{2Q_m^2}{4R^4+\l Q_m^2}\,,
\end{equation}
so that $T^{(\l)0}{}_0$ depends on $\l$ both explicitly, and through the radial function $R$ via \eqref{rofz}. With the substitution \eqref{rofz}, we find,  for the Hamiltonian density as a function of the radial variable $Z$:
\begin{equation}
\mathcal{H}^{(\l)}(Z) =  -T^{(\l)0}{}_0(R(Z))= \frac{1}{\l}\left(1-\sqrt{1-\frac{\lambda  Q_m^2}{Z^4}}\right)\,,
\end{equation}
which yields the expression of the density $\mathcal{H}^{(\l)}$ in the reference frame induced by $h^{(\l)}_{\mu\nu}$. 

Finally, to find the deformed solution in the original background metric $g_{\mu \nu}$, the geometric approach adopted in this paper includes a final formal step in which $h^{(\l)}_{\mu \nu}$ is renamed as $g_{\mu \nu}$. In this simple case, this corresponds to replacing  $Z$ with $R$, leading to equation ($M19$).
\end{document}